\documentclass[paper]{JHEP3}
\pdfoutput=1
\usepackage{amsmath,amssymb,amsthm,amscd,graphicx}
\input epsf.sty

\theoremstyle{definition}

\newcommand{\CM}{{\cal M}}
\newcommand{\CN}{{\cal N}}
\newcommand{\CO}{{\cal O}}

\newcommand{\CW}{{\cal W}}

\def\IZ{{\mathbb Z}}
\def\IR{{\mathbb R}}
\def\IC{{\mathbb C}}
\def\IP{{\mathbb P}}

\def\IS{{\mathbb S}}

\newcommand{\re}{{\rm e}}
\newcommand{\ri}{{\rm i}}
\newcommand{\rd}{{\rm d}}

\newcommand{\be}{\begin{equation}}
\newcommand{\ee}{\end{equation}}
\newcommand{\ba}{\begin{aligned}}
\newcommand{\ea}{\end{aligned}}
\newcommand{\ben}{\begin{eqnarray}\displaystyle}
\newcommand{\een}{\end{eqnarray}}

\newcommand{\sectiono}[1]{\section{#1}\setcounter{equation}{0}}

\newdimen\tableauside\tableauside=1.0ex
\newdimen\tableaurule\tableaurule=0.4pt
\newdimen\tableaustep
\def\phantomhrule#1{\hbox{\vbox to0pt{\hrule height\tableaurule width#1\vss}}}
\def\phantomvrule#1{\vbox{\hbox to0pt{\vrule width\tableaurule height#1\hss}}}
\def\sqr{\vbox{%
  \phantomhrule\tableaustep
  \hbox{\phantomvrule\tableaustep\kern\tableaustep\phantomvrule\tableaustep}%
  \hbox{\vbox{\phantomhrule\tableauside}\kern-\tableaurule}}}
\def\squares#1{\hbox{\count0=#1\noindent\loop\sqr
  \advance\count0 by-1 \ifnum\count0>0\repeat}}
\def\tableau#1{\vcenter{\offinterlineskip
  \tableaustep=\tableauside\advance\tableaustep by-\tableaurule
  \kern\normallineskip\hbox
    {\kern\normallineskip\vbox
      {\gettableau#1 0 }%
     \kern\normallineskip\kern\tableaurule}%
  \kern\normallineskip\kern\tableaurule}}
\def\gettableau#1{\ifnum#1=0\let\next=\null\else
\squares{#1}\let\next=\gettableau\fi\next}

\newcommand{\figref}[1]{Fig.~\protect\ref{#1}}

\title{Membrane instantons from a semiclassical TBA}

\author{
Flavio Calvo and Marcos Mari\~no
\\
D\'epartement de Physique Th\'eorique et Section de Math\'ematiques,\\
Universit\'e de Gen\`eve, Gen\`eve, CH-1211 Switzerland\\
\\
\email{marcos.marino@unige.ch}}

\abstract{The partition function on the three-sphere of ABJM 
theory contains non-perturbative corrections which correspond to membrane instantons in M-theory. These corrections can be studied in the Fermi gas approach to the partition function, and they are encoded in a system of integral equations of the TBA type. We study a semiclassical or WKB expansion of this TBA system in the ABJM coupling $k$, which corresponds to the strong coupling expansion of the type IIA string. This allows us to study membrane instanton corrections in M-theory at high order in the WKB expansion. Using these WKB results, we verify the conjectures for the form of the one-instanton correction at finite $k$ proposed recently by Hatsuda, Moriyama and Okuyama (HMO), which are in turn based on a conjectural cancellation of divergences between worldsheet instantons and membrane instantons. The HMO cancellation mechanism is important since it shows in a precise, quantitative way, that the perturbative genus expansion is radically insufficient at strong coupling, and that non-perturbative membrane effects are essential to make sense of the theory. We propose analytic expressions in $k$ for the full two-membrane instanton correction and for higher-order non-perturbative terms, which pass many consistency checks and provide further evidence for the HMO mechanism. 
}

\begin{document}

\sectiono{Introduction}

The partition functions on the three-sphere of ${\cal N}\ge 2$ Chern--Simons--matter theories can be reduced by localization to matrix models \cite{kwy, jafferis,hama} which have been much studied in the last years. Surprisingly, these matrix models contain an enormous amount of information. When studied in the large $N$ limit, this information can be decoded in terms of M-theory AdS duals. For example, the leading large $N$ free energy can be seen to reproduce the gravity action evaluated on-shell, as first found in \cite{dmp} in the case of ABJM theory \cite{abjm} (see \cite{lectures} for a review and a list of relevant references). 

 Although most of the work done on these models has focused on the leading order contribution at large $N$, from the point of view of M-theory and quantum gravity the most interesting information is contained in the subleading corrections. For example, the subleading logarithmic correction in $N$ corresponds to a one-loop correction in quantum supergravity, as it has been shown in \cite{sen}. There are two types of corrections in the large $N$ expansion: the perturbative corrections in $1/N$, and the non-perturbative or exponentially small corrections. The perturbative corrections can be computed in closed form in a large class of $\CN=3$ theories \cite{mp}, and they can be resummed in terms of an Airy function, as first shown in \cite{fhm} for ABJM theory. The non-perturbative corrections can be in turn divided in two types: the ones due to worldsheet instanton corrections in the AdS dual, and the ones due to more general membrane instanton corrections in M-theory \cite{bbs}. The appearance of worldsheet instanton corrections was anticipated in \cite{cagnazzo}, in the case of ABJM theory. They appear naturally in the 't Hooft expansion of the matrix model, at strong 't Hooft coupling, and were determined in a systematic, recursive way in \cite{dmp}, in a weak string coupling expansion. 
 
 The study of membrane instanton corrections is more challenging, since these are non-perturbative effects both in the string dual and in the large $N$ matrix model. In \cite{dmpnp} some information about these instantons (like their action) was obtained from the study of the large genus asymptotics of the 't Hooft expansion, but no concrete recipe was given to calculate them. In \cite{mp}, a new method was introduced to study the ABJM matrix model and its close cousins, based on an equivalence with a quantum Fermi gas. In the Fermi gas approach, the Planck constant is naturally identified with the inverse string coupling, and the semiclassical limit of the gas corresponds to the strong string coupling limit. One of the main virtues of the Fermi gas approach is that it makes possible to calculate membrane instanton effects systematically, at least in the WKB expansion. This opened the way for a quantitative determination of non-perturbative effects in the M-theory duals to Chern--Simons--matter theories. 
 
In a recent paper \cite{hmo2}, Hatsuda, Moriyama and Okuyama (HMO) made various crucial observations on the structure of non-perturbative corrections in ABJM theory, where the inverse string coupling is essentially given by the Chern--Simons level $k$. First of all, they noticed that the worldsheet instanton contributions to the free energy can be resummed at finite $k$ by using a Gopakumar--Vafa representation. The resulting expressions are divergent for integer $k$\footnote{This had been already noticed in 2011 in unpublished work by the second author and Pavel Putrov.}. Since the free energy is finite for any value of $k$, these divergences have to disappear in the final answer, and \cite{hmo2} suggested that they cancel against similar divergences in the contributions of membrane instantons, in such a way that the total sum of all non-perturbative effects at integer $k$ is finite. We will call this the HMO cancellation mechanism. This mechanism is beautiful and natural, and we believe is of deep conceptual importance for the understanding of M-theory. It shows, in a precise and quantitative way, that the genus expansion based on strings is essentially meaningless: in the non-perturbative completion of type IIA string theory at finite, integer $k$ through M-theory, only the combination of membrane instantons and worldsheet instantons makes sense. 
 
 In some cases, the HMO mechanism gives a set of constraints for the membrane instanton corrections at finite $k$. Using these constraints, as well as the first few terms of the semiclassical expansion at small $k$ obtained in \cite{mp}, an expression for the one-instanton membrane correction was proposed in \cite{hmo2} which passes many consistency checks. It reproduces for example the low order, non-perturbative corrections to the free energy, for small integer values of $k$. 
 
 The purpose of the present note is twofold. It was pointed out in \cite{mp} that the Hamiltonian problem appearing in the Fermi gas approach to ABJM theory can be studied with a pair of TBA equations first considered in \cite{cv,fs} and studied in detail in \cite{zamo}. This observation was exploited in \cite{hmo,py,hmo2}, where the TBA equations were solved for finite $k$ but small chemical potential. This makes it possible to compute the partition function of ABJM theory for finite $k$ and small $N$. On the other hand, the Fermi gas can be studied in the WKB approximation, at small $k$, and this provided many valuable insights into the problem \cite{mp}. Our first goal in this paper is to develop a semiclassical or WKB expansion directly in the TBA equations. This leads to an algorithm which calculates the 
 grand potential of the ABJM model systematically, as a power series in $k$, and arbitrary chemical potential. This method is more powerful than the original WKB expansion of the Fermi gas studied originally in \cite{mp}, and one can easily push the computation to higher orders. 
 
 Our second goal is to use this information to further explore the cancellation mechanism proposed in \cite{hmo2}. We verify that the expression for the membrane one-instanton correction proposed in \cite{hmo2} agrees with the WKB expansion to high order, and we propose analytic expressions for 
 the full two-instanton correction and for some higher-order terms in the non-perturbative expansion. Our proposal for the two-instanton correction passes all the consistency tests, and in particular agrees with the results for the grand potential at $k=1,3$ obtained in \cite{hmo2}. Our WKB results confirm then the HMO cancellation mechanism and are very helpful in obtaining conjectural expressions at finite $k$. 
 
 The organization of the paper is as follows. In section 2 we review some general results on non-perturbative effects in ABJM theory. In section 3 we explain how the Fermi gas approach of \cite{mp} can be reformulated in terms of the TBA system considered by Al. Zamolodchikov in \cite{zamo}. In section 4 we study in detail the WKB expansion of the TBA system and explain how it leads to the semiclassical expansion for the grand potential considered in \cite{mp}. In section 5 we give an application of the WKB expansion to the calculation of membrane instantons: we test the conjecture for one-membrane instanton corrections in \cite{hmo2}, and we propose an exact formula for the two-membrane instanton contributions. Finally, we conclude with some open problems. In the first appendix we list some useful results for the integration of generalized hypergeometric functions, and in the second appendix we list WKB expansions for some low order contributions of 
 membrane instantons.

\sectiono{Perturbative and non-perturbative aspects of ABJM theory}

The quantity we will focus on in this paper is the partition function of ABJM theory on the three-sphere, $Z(N,k)$, which is given by the matrix integral \cite{kwy}  
\be
\label{abjmmatrix}
\ba
&Z_{\rm ABJM}(N)\\
&={1\over N!^2} \int {\rd ^N \mu \over (2\pi)^N} {\rd ^N \nu \over (2\pi)^N} {\prod_{i<j} \left[ 2 \sinh \left( {\mu_i -\mu_j \over 2} \right)\right]^2
  \left[ 2 \sinh \left( {\nu_i -\nu_j \over 2} \right)\right]^2 \over \prod_{i,j} \left[ 2 \cosh \left( {\mu_i -\nu_j \over 2} \right)\right]^2 } 
  \exp \left[ {\ri k \over 4 \pi} \sum_{i=1}^N (\mu_i^2 -\nu_i^2) \right].
  \ea
  \ee
This matrix integral can be calculated in the 't Hooft expansion
\be
\label{thooftl}
N \rightarrow \infty, \quad \lambda={N\over k}\, \,\,  \text{fixed}
\ee
by using techniques of matrix model theory and topological string theory \cite{mp-exact,dmp}. 
In particular, one can obtain explicit formulae for the genus $g$ free energies appearing in the $1/N$ expansion
\be
F(\lambda,g_s)=\sum_{g=0}^{\infty} g_s^{2g-2} F_g(\lambda),
\ee
where 
\be
g_s ={2 \pi \ri \over k}.
\ee
The genus $g$ free energies $F_g(\lambda)$ obtained in this way are exact interpolating functions, and they can be studied in various regimes of the 
't Hooft coupling. When expanded at strong coupling, they have the structure
\be
F_g(\lambda)=F_g^{\rm p}( \lambda) + F_g^{\rm np}(  \lambda). 
\ee
The first term represents the perturbative contribution in $\alpha'$, while the second term is non-perturbative in $\alpha'$, 
\be
\label{winst}
F_g^{\rm np}( \lambda) \sim \CO\left(\re^{-2 \pi {\sqrt{2\lambda}}}\right).
\ee
The type IIA dual of ABJM theory involves the space AdS$_4\times \IC\IP^3$ \cite{abjm}, and this geometry 
supports worldsheet instantons wrapping a $\IC\IP^1 \subset \IC\IP^3$ \cite{cagnazzo}. The non-perturbative piece (\ref{winst}) was interpreted in \cite{dmp} as the contribution of these 
worldsheet instantons. 

Besides the non-perturbative effects in $\alpha'$, one can use the connection between the large-order behavior of perturbation theory and instantons to deduce the 
structure of non-perturbative effects in the string coupling constant. In \cite{dmpnp} a detailed analysis showed that these effects would have the form 
\be
\label{membranea}
\exp\left( -k \pi {\sqrt{2 \lambda}} \right) 
\ee
at large $\lambda$. It was also proposed in \cite{dmpnp} that the source of these effects are D2-branes wrapped around three-cycles in the target space. An appropriate, explicit family of generalized Lagrangian submanifolds with the topology of $\IR\IP^3\subset \IC\IP^3$ was proposed as an explicit candidate for these cycles. We will refer to these non-perturbative effects as membrane instanton effects: they can be interpreted as M2-instantons in M-theory \cite{bbs}, where the M2-brane wraps a three-cycle inside $\IS^7/\IZ_k$ which is the lift of the three-cycle in $\IC\IP^3$. Notice that these membrane instanton effects are invisible in ordinary string perturbation theory.  

Further information on the membrane intantons can be obtained by using the Fermi gas approach introduced in \cite{mp}. In this approach, one first notices (see also \cite{kwytwo}) that the matrix integral (\ref{abjmmatrix}) can be written as 
\be
\label{fg-matrix}
Z(N,k)={1 \over N!} \sum_{\sigma  \in S_N} (-1)^{\epsilon(\sigma)}  \int  {\rd ^N x \over (2 \pi k)^N} {1\over  \prod_{i} 2 \cosh\left(  {x_i  \over 2}  \right)
2 \cosh\left( {x_i - x_{\sigma(i)} \over 2 k} \right)}.
\ee
This in turn can be interpreted as the canonical partition function of a one-dimensional Fermi gas with a non-trivial one-particle density matrix
\be
\label{densitymat}
\rho(x_1, x_2)={1\over 2 \pi k} {1\over \left( 2 \cosh  {x_1 \over 2}  \right)^{1/2} }  {1\over \left( 2 \cosh {x_2  \over 2} \right)^{1/2} } {1\over 
2 \cosh\left( {x_1 - x_2\over 2 k} \right)}.
\ee
The one-particle Hamiltonian $\hat H$ of this system is then defined as
\be
\label{onepH}
\hat \rho=\re^{-\hat H}, \qquad \langle x_1 | \hat \rho | x_2 \rangle = \rho(x_1, x_2), 
\ee
and the Planck constant is 
\be
\hbar = 2\pi k. 
\ee
The semiclassical or WKB expansion is then around $k=0$, and it corresponds to the strong string coupling expansion in the type IIA dual. 

The Fermi gas approach suggests to look instead to the grand partition function (see also \cite{okuyama})
\be
\label{grand}
\Xi(\mu, k)=1+\sum_{N=1}^\infty Z(N,k) z^N, 
\ee
where 
\be
z=\re^{\mu}
\ee
plays the r\^ole of the fugacity and $\mu$ is the chemical potential. The grand potential is then defined as
\be
J(\mu,k) =\log \Xi (\mu, k). 
\ee
The canonical partition function is recovered from the grand-canonical potential as
\be
\label{exactinverse}
Z(N, k) =\oint {\rd z \over 2 \pi \ri } {\Xi (\mu, k) \over z^{N+1}}.
\ee
At large $N$, this integral can be computed by applying the saddle-point method to
\be
\label{muint}
Z(N,k) ={1\over 2 \pi \ri} \int \rd \mu \, \exp\left[J(\mu,k) - \mu N\right].
\ee

As shown in \cite{mp}, the grand potential is the sum of a perturbative and a non-perturbative piece, 
\be
J(\mu, k)=J^{\rm p}(\mu, k) + J^{\rm np}(\mu, k). 
\ee
The perturbative piece is a cubic polynomial in $\mu$: 
\be
J^{\rm p}(\mu,k)= {C(k) \over 3} \mu^3 + B(k) \mu + A(k).
\ee
The coefficients $C(k)$, $B(k)$ where computed in \cite{mp} for ABJM theory (in fact, the analogs of these coefficients for other 
$\CN=3$ theories can be also computed in closed form). The coefficient $A(k)$ can be computed in a WKB expansion around $k=0$ \cite{mp}, and the all-orders result was 
conjectured in \cite{hanada}. In this paper we will be interested in the 
non-perturbative piece $J^{\rm np}(\mu,k)$. In general one has the following result: 
\be
\label{np}
J^{\rm np}(\mu, k)= J^{\rm M2}(\mu, k) + J^{\rm WS}(\mu, k) + \cdots. 
\ee
Let us explain this structure in more detail. The first term $J^{\rm M2}(\mu, k)$ has the following expansion for $\mu \gg 1$, 
\be
\label{gen-M2}
J^{\rm M2}(\mu, k) = \sum_{\ell\ge 1} \left(a_\ell (k) \mu^2 + b_\ell (k)  \mu + c_\ell(k) \right) {\rm e}^{-2 \ell \mu}. 
\ee
This type of contributions to the grand potential was first found in \cite{mp} and it was interpreted there as due to membrane instantons, i.e. to 
M2-branes wrapping a three-cycle $\CM \subset \IS^7/\IZ_k$ which is a lift of a three-cycle in $\IC\IP^3$. The positive integer $\ell$ is the winding number of the wrapping. We will refer 
to the $\ell$-th term in the infinite series (\ref{gen-M2}) as the the contribution of the $\ell$-membrane instanton. The coefficients $a_\ell(k)$, $b_\ell(k)$ and $c_\ell(k)$ are non-trivial functions of $k$. Using the WKB expansion of the quantum Fermi gas, it was shown in \cite{mp} that they have a perturbative expansion around $k=0$ of the form 
\be
a_\ell(k)= {1\over k} \sum_{n=0}^\infty a_{\ell,n} k^{2n}. 
\ee
A similar expansion holds for $b_\ell(k)$ and $c_\ell(k)$, with coefficients $b_{\ell,n}$, $c_{\ell,n}$, respectively. The WKB expansion was obtained in \cite{mp} up to order $n=2$. One goal of this paper will be to obtain a more efficient method to calculate the WKB expansion. 

The second term in (\ref{np}), $J^{\rm WS}(\mu, k)$, is the contribution of worldsheet instantons wrapping $\IC\IP^1 \subset \IC\IP^3$. It has the following expansion for $\mu \gg 1$, 
\be
J^{\rm WS}(\mu, k)= \sum_{m=1}^{\infty} d_m (k) \re^{-{4 m \mu \over k}}. 
\ee
In \cite{hmo2} a very useful formula was proposed for $J^{\rm WS}(\mu, k)$: by using the fact \cite{mp-exact,dmp} that the ABJM matrix integral is dual to topological string theory on local $\IP^1 \times \IP^1$, 
one can write
\be
\label{gvone}
J^{\rm WS}(\mu, k)= \sum_{n,g,d\ge1} n_d^g \left( \sin {2 \pi n \over k} \right)^{2g-2} {(-1)^{dn}  \over n}{\rm e}^{-{4 dn \mu \over k}}.
\ee
In this formula, $n_d^g$ is the weighted sum of the Gopakumar--Vafa invariants \cite{gv} of local $\IP^1 \times \IP^1$, 
\be
\label{gvtwo}
n_d^g= \sum_{d_1+d_2 =d} n_{d_1, d_2}^g. 
\ee

\FIGURE{
\includegraphics[height=6cm]{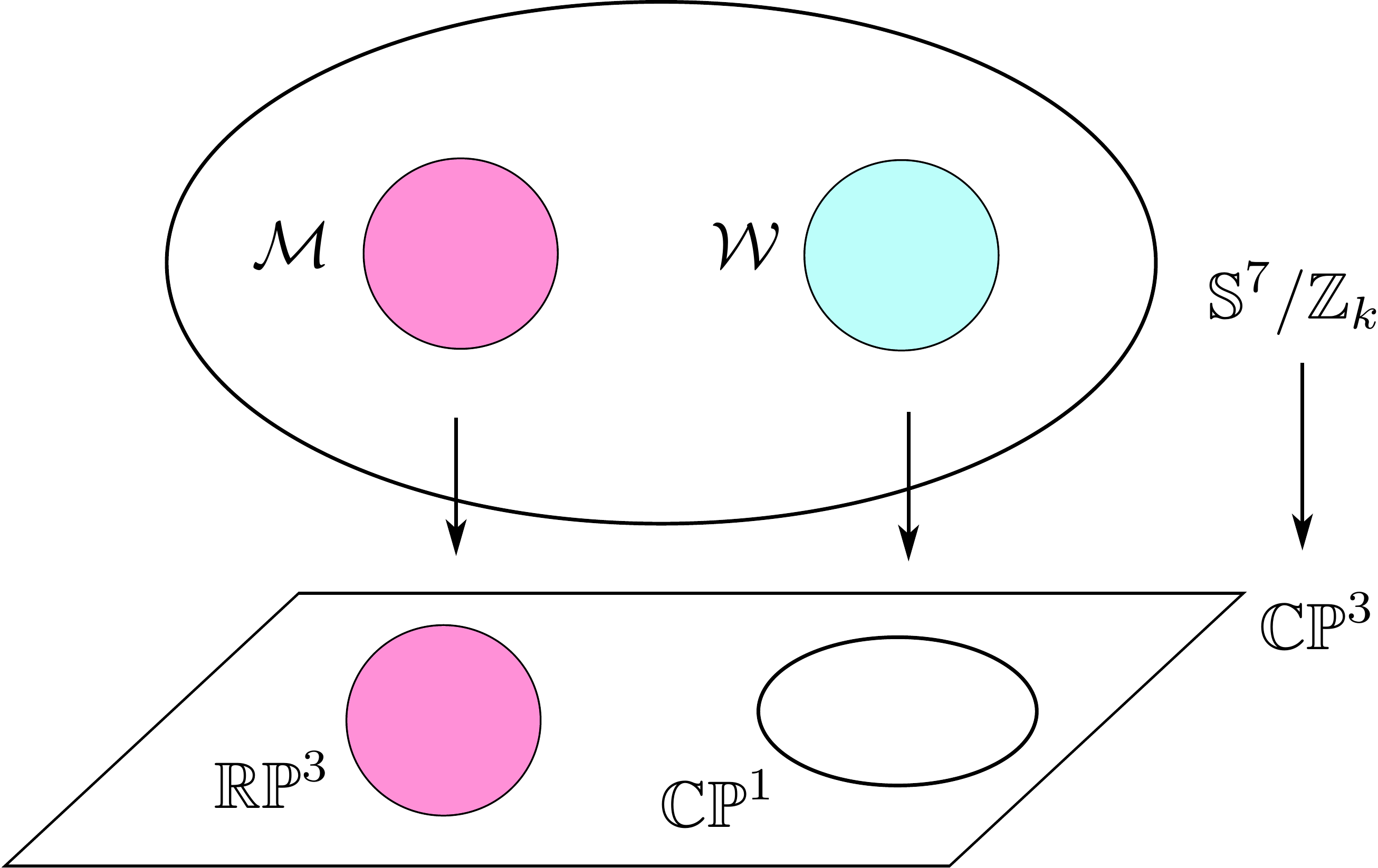} 
\caption{From the M-theory point of view, there are two types of non-perturbative effects in ABJM theory: M2-branes can wrap a cycle $\CM \subset \IS^7/\IZ_k$ which descends to an $\IR\IP^3 \subset \IC\IP^3$ cycle in the type IIA target; or they can wrap a cycle $\CW$ which descends to $\IC\IP^1\subset \IC\IP^3$. The most general M2-brane configuration wraps $\ell$ times the cycle $\CM$ and $m$ times the cycle $\CW$.}
\label{in-fig}
}

Notice that, from the point of view of M-theory, both types of instantons are due to M2-branes wrapping three-cycles: membrane instantons correspond to a three-cycle $\CM$ which descends to a three-cycle wrapped by D2-branes in the type IIA target (if the proposal of \cite{dmpnp} is correct, this three-cycle is an $\IR\IP^3 \subset \IC\IP^3$); worldsheet instantons correspond to M2-branes wrapping the three-cycle $\CW=\IS^3/\IZ_k$, which descends to $\IC\IP^1\subset \IC\IP^3$. Therefore, from the point of view of M-theory, there are {\it two} types of three-cycles, $\CM$ and $\CW$, and the most general M2-brane configuration wraps $\ell$ times the cycle $\CM$ and $m$ times the cycle $\CW$, see \figref{in-fig}. We should then expect that the most general contribution to $J^{\rm np}(\mu)$ is of the form 
\be
\label{bound-state}
J^{\rm np}(\mu,k)= \sum_{\ell,m\ge 1} f_{\ell,m}(k, \mu) \exp\left[ -\left( 2\ell + {4 m \over k}  \right) \mu \right].
\ee
The contribution $J^{\rm M2}(\mu, k)$ corresponds to $m=0$, while the contribution $J^{\rm WS}(\mu, k)$ corresponds to $\ell=0$. The existence of such bound states, labeled by two integers $(\ell, m)$, was first indicated in \cite{hmo2} in order to interpret their data on the grand potential. From the point of view of M-theory they are completely natural, since they correspond to the most general set of supersymmetric cycles in the geometry. 

Can we compute the contribution of these non-perturbative states? In the case of worldsheet instantons, the expressions (\ref{gvone}) and (\ref{gvtwo}) reduce the problem to the determination of 
the Gopakumar--Vafa invariants of local $\IP^1 \times \IP^1$. This can be done in many different ways, and there is no difficulty in calculating $J^{\rm WS}(\mu, k)$ to any desired order. The calculation of $J^{\rm M2}(\mu, k)$ can be done order by order in $k$ and at all possible orders in the membrane instanton number by using the WKB method of \cite{mp}. In the remaining of the paper, we will 
develop a more efficient method, based on the TBA equations of \cite{zamo}, to calculate this perturbative expansion around $k=0$.  However, there is no known procedure to determine the contribution of bound states appearing in (\ref{bound-state}), not even at small $k$. This is probably the most important open problem in this subject, as we will mention in the final section of the paper.

\sectiono{The Fermi gas/TBA approach to ABJM theory}

\subsection{Fredholm determinants and TBA}
We will now summarize some results from \cite{zamo,tw} which are needed in this paper. 

In \cite{zamo}, the following integral kernel is considered
\be
\label{kernel}
K(\theta, \theta')={1\over 2 \pi} {\exp\left(-u(\theta) - u (\theta')\right) \over 2 \cosh{ \theta-\theta'\over 2}}. 
\ee
This defines a homogeneous Fredholm integral equation of the second kind, of the form 
\be
\label{eigen-K}
\int_{-\infty}^{\infty} K(\theta, \theta') f(\theta') = \lambda f(\theta), 
\ee
Let $\lambda_a$ be the possible eigenvalues appearing in (\ref{eigen-K}), and let us introduce the Fredholm determinant
\be
\label{fredholm}
\Xi (z) =\prod_n (1+ z \lambda_a). 
\ee
We will regard $\Xi$ as a grand canonical partition function. The grand potential is then given by 
\be
J(z)=\log \, \Xi(z)
\ee
and it has the expansion 
\be
 J(z)=-\sum_{\ell=1}^{\infty} {(-z)^\ell \over \ell} Z_\ell
 \ee
where $Z_\ell$ is given by the integral
 \be
 Z_{\ell}=\int_{-\infty}^{\infty} \prod_{i=1}^\ell {\re^{-2u (\theta_i)} \over 2 \cosh {\theta_i-\theta_{i+1} \over 2}}{\rd \theta_i \over 2 \pi}
 \ee
 with the periodicity condition 
 \be
 \theta_{\ell+1}=\theta_1. 
 \ee

We now introduce the iterated integral 
 \be
 \label{it-int}
 R_\ell (\theta)=\re^{-2 u(\theta)} \int_{-\infty}^{\infty} {\re^{-2u (\theta_1)-\cdots -2u(\theta_\ell) } \over  \cosh {\theta-\theta_{1} \over 2}   \cosh {\theta_1-\theta_2 \over 2} \cdots 
  \cosh {\theta_\ell-\theta \over 2}}\rd \theta_1 \cdots \rd \theta_\ell, \qquad \ell \ge 1,
 \ee
 and
 \be
 R_0(\theta)=\re^{-2 u(\theta)}.
 \ee
Notice that 
\be
\label{RZ}
\int_{-\infty}^{\infty} \rd \theta \, R_\ell(\theta)= (4 \pi)^{\ell+1} Z_{\ell+1} 
\ee
and the generating series 
\be
R(\theta|z)= \sum_{\ell \ge 0} \left( -{z \over 4 \pi} \right)^\ell R_\ell (\theta)
\ee
satisfies
\be
\label{zamoJ}
\int_{-\infty}^{\infty}  {\rd \theta \over 4 \pi} R(\theta|z)=\sum_{\ell \ge 0} \left( -z\right)^\ell Z_{\ell+1} ={\partial J \over \partial z}.
\ee

It was conjectured in \cite{zamo} and proved in \cite{tw} that the function $R(\theta|z)$ can be obtained by using TBA-like equations which first appeared in the context of two-dimensional $\CN=2$ theories \cite{cv,fs}. We first define
\be
\label{rs}
\ba
R_+(\theta|z)&={1\over 2} \left( R(\theta|z)+ R(\theta|-z) \right), \\
R_-(\theta|z)&={1\over 2} \left( R(\theta|z)- R(\theta|-z) \right).
\ea
\ee
Let us now consider the TBA system
\be
\label{tba}
\ba
2u(\theta)&= \epsilon(\theta) + \int_{-\infty}^{\infty} {\rd \theta' \over 2\pi} {\log \left(1+\eta^2(\theta')\right) \over \cosh(\theta-\theta')}, \\
\eta(\theta)&=-z \int_{-\infty}^{\infty} {\rd \theta'  \over  2\pi} {\re^{-\epsilon(\theta')}\over \cosh(\theta-\theta')}. 
\ea
\ee
Then, one has that
\be
\ba
R_+(\theta|z)&= \re^{-\epsilon(\theta)},\\
R_-(\theta|z)&=R_+(\theta|z) \int_{-\infty}^{\infty} {\rd \theta' \over \pi} {\arctan\, \eta(\theta') \over \cosh^2(\theta-\theta')}.
\ea
\ee
This conjecture has been proved in \cite{tw} for general $u(\theta)$. In general, the system (\ref{tba}) has to be solved numerically, although an exact solution exists for $u(\theta) =\re^{\theta}$ in terms of Airy functions \cite{fendley}.  

\subsection{From the ABJM Fermi gas to TBA}

Let us now find the relation between the Fermi gas approach to ABJM theory and the TBA system considered above (see also \cite{py,hmo}). We start from the expression (\ref{fg-matrix}) for the partition function. We can now regard the density matrix as an integral kernel with the structure of (\ref{kernel}). The Fredholm determinant of this kernel is nothing but the grand partition function of the Fermi gas. We can then use the results of \cite{zamo,tw} to write a TBA-like equation determining its grand potential. 
Let us consider the quantities
\be
\rho^{\ell +1}(x,x)= \langle x | \rho^{\ell +1} |x \rangle={\re^{-2\upsilon (x)} \over 2 \pi k} \int_{-\infty}^{\infty} {\rd x_1 \cdots \rd x_\ell \over (2 \pi k)^\ell}  {\re^{-2\upsilon(x_1)} \cdots \re^{-2\upsilon(x_\ell)} \over   2 \cosh\left( {x - x_1 \over 2 k} \right) \cdots 
 2 \cosh\left( {x_\ell - x\over 2 k} \right)}
 \ee
 where 
\be
\upsilon(x)={1\over2} \log \left(2 \cosh{ x \over 2} \right). 
\ee
If we now change variables
\be
x_i=k \theta_i.
\ee
and compare with (\ref{it-int}), we find 
\be
\rho^{\ell +1}(x,x)= {1\over (4 \pi)^{\ell+1} k } R_\ell(x)
 \ee
where we have denoted 
\be
R_\ell (x)\equiv R_\ell \left( \theta ={x\over k} \right).
\ee
The function $R_\ell(\theta)$ is calculated with the TBA system (\ref{tba}) and the potential 
\be
\label{utheta}
u(\theta)={1\over2} \log \left(2 \cosh{ k \theta \over 2} \right)
\ee
which depends explicitly on $k$. The grand potential is given by 
\be
\label{grand-k}
{\partial J \over \partial z}= {1\over 4 \pi k} \int_{-\infty}^{\infty} \rd x R\left( x |z\right). 
\ee

Notice that the function $R(x|z)$ can be written as 
\be
\label{rxz}
R\left( x |z\right)= {4 \pi k \over z} \left \langle x \left| {1\over \re^{\hat H-\mu} +1}\right| x \right\rangle.
\ee
The quantity appearing in (\ref{rxz}) is essentially the full quantum-corrected version of the 
Fermi momentum $p_F(x)$ of the Fermi gas. Indeed, (\ref{rxz}) can be computed by using the Wigner map, which associates to any quantum operator $\CO$ 
a function $\CO_{\rm W}(x,p)$ in phase 
space (see \cite{mp} for details):
\be
\label{r-fg}
R\left( x \right|z)= {1\over  \pi z} \int_{-\infty}^{\infty} \rd p \left( {1\over \re^{\hat H-\mu} +1} \right)_{\rm W}={2  \over \pi z} p_F(x).
\ee

There is an important property of the TBA system of \cite{zamo} which is worth discussing in some detail. Notice that the functions $\epsilon(\theta)$, $\eta(\theta)$ make it possible to calculate {\it both} $R(x|z)$ and $R(x|-z)$. The last quantity is given by  
\be
R\left( x|-z\right)= {4 \pi k \over z}  \left \langle x \left| {1\over \re^{\hat H-\mu} -1}\right| x \right\rangle, 
\ee
and it corresponds to the same one-particle Hamiltonian (\ref{onepH}) but with {\it Bose--Einstein} statistics. If we now take into account the expression (\ref{fredholm}), we deduce that for Bose--Einstein statistics there is a physical singularity at 
\be
z=\lambda_0^{-1}>0
\ee
where $\lambda_0$ is the largest eigenvalue of the non-negative Hilbert--Schmidt operator $\rho(x_1, x_2)$. This singularity corresponds of course to the onset of Bose--Einstein condensation in the gas, and as a consequence the functions $R_\pm (x|z)$ will have singularities in the $x$-plane for $z\ge  \lambda_0^{-1}$. But this is precisely the regime in which we are interested, since large $N $ corresponds to $\mu \gg 1$. Of course, the singularity at large positive $z$ {\it cancels} once one adds up $R_+$ and $R_-$. 

The appearance of this Bose--Einstein condensate has prevented the direct study of the large $N$ limit of ABJM theory with the TBA equations (\ref{tba}), even numerically: the standard iteration of the integral equations does not converge when $z$ is large enough. In the papers \cite{hmo,py} they study in fact the 
small $z$ regime of (\ref{tba}) at finite $k$ in order to extract the small $z$ expansion of the grand partition function and therefore the canonical partition functions $Z(N,k)$ for small $N$. The large $N$ limit is then extracted from these small $N$ results by numerical extrapolation \cite{py,hmo2}. We will now propose another approach to study the TBA equations.

\sectiono{The semiclassical TBA system} 

\subsection{General aspects}
As we showed in the last section, the TBA system is equivalent to the Fermi gas picture. In particular, after setting $\theta=x /k$, the function $R(x|z)$ is the Fermi momentum of the gas. As shown in \cite{mp}, one can do a systematic computation of all quantities in the Fermi gas in a WKB expansion, i.e. in a perturbative expansion in $k$. This suggests studying the TBA equation in a scaling regime where $k$ is small and one can construct a perturbative expansion around $k=0$. To do this, we set $\theta=x /k$ and write (\ref{tba}) as
\be
\label{tba-k}
\ba
U(x)&= \epsilon\left(x\right) + \int_{-\infty}^{\infty} {\rd x' \over 2\pi k } {\log \left(1+\eta^2(x')\right) \over \cosh\left({x-x' \over k} \right)}, \\
\eta(x)&=-z \int_{-\infty}^{\infty} {\rd x'  \over  2\pi k } {\re^{-\epsilon(x')}\over \cosh\left({x-x' \over k} \right)}. 
\ea
\ee
where
\be
\label{Uu}
U(x)=2 u \left( \theta={x \over k} \right). 
\ee
The second equation in (\ref{rs}) becomes
\be
\label{rsx}
R_-(x|z)=R_+(x |z) \int_{-\infty}^{\infty} {\rd x'  \over \pi k} \, {\arctan\, \eta(x') \over   \cosh^2\left({x - x'\over k} \right)}.
\ee

In the case of ABJM theory, where the potential is given by (\ref{utheta}), we have 
\be
U(x)= \log \left( 2 \cosh {x\over 2} \right)
\ee
and it is independent of $k$. Notice that in \cite{zamo,tw} one considers a general potential $u(\theta)$, with no $k$ parameter to start with, 
but we can study a one-parameter deformation of the problem by considering a potential $u(k \theta)$, in such a way that $U(x)$ in (\ref{Uu}) is independent of $k$. The original problem is then obtained when $k=1$.

The advantage of the $k$-dependent equations (\ref{tba-k}), (\ref{rsx}) is that they admit a systematic, perturbative expansion around $k=0$, where they can be solved algebraically. This can be seen from the fact that, when $k\rightarrow 0$, the kernel becomes a $\delta$-function: 
\be
\lim_{k \to 0} {1\over 2 \pi k \cosh \left({x\over k}\right)} = {1\over 2} \delta(x),
\ee
and the integral equations become algebraic equations. The most convenient form of the TBA system for the small $k$ expansion involves finite difference equations, and it was already considered (for $k=1$) in \cite{zamo} and specially in \cite{tw}. In more complicated models, the system of TBA integral equations, when written in terms of difference equations, is usually called the functional $Y$ system. This form of the TBA system can be easily obtained by Fourier transform. If we denote
\be
F[f(x);p]=\hat f(p)= {1\over {\sqrt{2\pi}}} \int_{-\infty}^{\infty} \rd x f(x) \re^{\ri p x}, 
\ee
we obtain, from the first equation in (\ref{tba-k}), 
\be
\label{ft} 
2 \cosh\left( {\pi k p \over 2}\right) \left( \widehat U(p)-\widehat \epsilon (p) \right) = F\left[ \log (1+ \eta^2(x)); p\right].
\ee
The first factor in the l.h.s. of (\ref{ft}) is the displacement operator
\be
\re^{{\ri k \pi \over 2} {\rd \over \rd x}}+ \re^{-{\ri k \pi \over 2} {\rd \over \rd x}}. 
\ee
Doing the same thing in the second equation of (\ref{tba-k}) we find that the two integral equations are equivalent (provided some mild analyticity conditions are satisfied, see \cite{tw}) to the two difference equations, 
\be
\ba
\epsilon \left( x +{\pi \ri k \over 2}\right) +\epsilon \left( x -{\pi \ri k \over 2}\right) &= U \left( x +{\pi \ri k \over 2}\right) +U \left( x -{\pi \ri k \over 2}\right) -\log (1+ \eta^2(x)),\\
\eta \left( x +{\pi \ri k \over 2}\right) +\eta \left( x -{\pi \ri k \over 2}\right) &= -z \re^{-\epsilon (x)}. 
\ea
\ee
Equivalently, in terms of $R_+(x)$, we have
\be
\label{semitba}
\ba
1+ \eta^2(x) &= R_+ \left( x +{\pi \ri k \over 2}\right) R_+ \left( x -{\pi \ri k \over 2}\right) \exp\left\{ U \left( x +{\pi \ri k \over 2}\right) +U \left( x -{\pi \ri k \over 2}\right) \right\},\\
-z R_+(x)&=  \eta \left( x +{\pi \ri k \over 2}\right) +\eta \left( x -{\pi \ri k \over 2}\right). 
\ea
\ee
The equation (\ref{rsx}) giving $R_-$ reads now, 
\be
\label{rsx-bis}
{R_- \left( x +{\pi \ri k \over 2}\right) \over R_+ \left( x +{\pi \ri k \over 2}\right)}- {R_- \left( x -{\pi \ri k \over 2}\right) \over R_+ \left( x -{\pi \ri k \over 2}\right)}=2 \ri k {\eta'(x) \over 1+ \eta^2(x)}.
\ee
When $k=1$, these equations have been written down in \cite{tw}. 

The TBA equations, in the form (\ref{semitba}), (\ref{rsx-bis}), can be solved systematically in an expansion around $k=0$. Let us denote 
\be
r(x)=R_+(x), \qquad t(x) = {R_-(x) \over R_+(x)}. 
\ee
We introduce the ansatz
\be
\ba
r(x) &=\sum_{n=0}^\infty r_n(x) k^{2n},\\
\eta(x)&=\sum_{n=0}^\infty \eta_n(x) k^{2n}, \\
t(x)&=\sum_{n=0}^\infty t_n(x) k^{2n}.
\ea
\ee
We then make a Taylor expansion in $k$ of the displaced functions in (\ref{semitba}), and we solve 
order by order in $k$ for $r_n$, $\eta_n$. Once $\eta$ is known, $t(x)$ can be obtained from  
\be
t(x) = {2 \over \pi} { \zeta  \over \sin  \zeta} \tan^{-1}(\eta), \qquad \zeta = {k \pi \over 2} \partial. 
\ee
The solution for these functions at leading order in the $k$ expansion is immediate, 
\be
\ba
r_0&= {2\over z} {\xi \over {\sqrt{1-\xi^2}}}, \\
\eta_0&=-{\xi \over {\sqrt{1-\xi^2}}},\\
t_0&=-{2\over \pi} \sin^{-1}(\xi), 
\ea
\ee
where we have introduced the variable
\be
\xi = {z\over 2} \re^{-U}. 
\ee
It is easy to see that, at each order in $k^{2n}$, we obtain from (\ref{semitba}) two {\it linear} equations for $r_n$ and $\eta_n$, which are solved in terms of derivatives of lower $r_{n'}$, $\eta_{n'}$ with $n'<n$ and of the potential $U(x)$. The procedure can be easily automatized in a computer code to obtain the functions $r_n, \eta_n$ to any desired order. 

It is now straightforward to write down a power series expansion for $R_-(x)$, $R(x)$. If we denote the coefficient of $k^{2n}$ in this expansion as $R_{-,n}(x)$, $R_n(x)$, we find
\be
\label{rss}
\ba
R_{-,n}(x)&=r_n(x)  t_0(x)+ B_n(x), \\
R_{n}(x)&=r_n(x)  \left(1+t_0(x) \right)+ B_n(x), 
\ea
\ee
where
\be
B_n(\xi)= \sum_{m=0}^{n-1} r_m (\xi)  t_{n-m}(\xi).
\ee
At leading order we find, for example, 
\be
\label{R0}
R_0=r_0\left(1+ t_0\right) ={4 \over \pi z} {\xi \over {\sqrt{1-\xi^2}}} \arccos(\xi).
\ee
Notice that both $r_0(x)$ and $\eta_0(x)$ have a branch cut, as functions of $\xi$, along $[1, \infty)$. However, $R_0(\xi)$ is holomorphic on the half-plane ${\rm Re}(\xi) >-1$, as it can be easily seen 
from the representation 
\be
R_0(\xi)= {4\over \pi  z} {\xi \over  {\sqrt{\xi^2-1}}} \log\left( \xi + {\sqrt{\xi^2-1}} \right). 
\ee
This is the semiclassical manifestation of the phenomenon we pointed out at the end of the previous section: the functions $\eta$, $R_+$ have a branch cut for 
\be
z> 2 \exp\left( {\rm min}\,  U \right), 
\ee
due to the appearance of a Bose--Einstein condensate. However, this branch cut cancels in the function $R_0$, since this function corresponds to Fermi statistics and there is no possible source of singularities for positive $z$ in that case.  

We can test the result for $R_0$ with a semiclassical calculation of (\ref{r-fg}) in the Fermi gas. At leading order in the WKB expansion, we have 
\be
 \left( {1\over \re^{\hat H-\mu} +1} \right)_{\rm W}\approx {1\over \re^{H(x,p)-\mu} +1},
 \ee
 where
 \be
 H(x,p) = \log \left( 2 \cosh {p \over 2}\right) + U(x), 
 \ee
 and 
\be
R_0= {1\over \pi z} \int_{-\infty}^{\infty} \rd p {1\over \xi^{-1} \cosh\left({p \over 2} \right) +1},
\ee
which indeed can be explicitly computed and it agrees with (\ref{R0})

\subsection{Semiclassical TBA for ABJM theory}

So far, we have considered the semiclassical TBA equations for a general potential $U(x)$. In the case of ABJM theory, we have 
\be
 \exp\left\{ U \left( x +{\pi \ri k \over 2}\right) +U \left( x -{\pi \ri k \over 2}\right) \right\}= {z^2 \over 4 \xi^2}- 4\sin^2 \left( {k \pi \over 4}\right)
 \ee
and it is useful to express everything in terms of $\xi$. It is then straightforward to study the semiclassical TBA system at higher order. For example, at next-to-leading order we find the equations
\be
\ba
2 \eta_0 \eta_1 &= {z^2 \over 4 \xi^2} \left[ {\pi^2 \over 4} \left( (r_0')^2 - r_0 r_0''\right) + 2 r_0 r_1 \right] -{\pi^2 r_0^2\over 4}, \\
- z r_1&= 2\eta_1 -{\pi^2 \over 4} \eta_0''
\ea
\ee
with the solution 
\be
\ba
r_1(\xi)&=\frac{\pi ^2 \xi ^3 \left(z^2 \left(2 \xi ^2+3\right)-80 \xi ^2\right)}{16 z^3 \left(1-\xi ^2\right)^{7/2}}, \\
\eta_1(\xi)&=-\frac{\pi ^2 \xi  \left(z^2 \left(4 \xi ^2+1\right)+16 \xi ^2 \left(\xi ^4-4 \xi ^2-2\right)\right)}{32 z^2 \left(1-\xi ^2\right)^{7/2}}.
\ea
\ee
It is easy to see from the recursion that the functions 
\be
(1-\xi^2) ^{3n+1/2} r_n (\xi), \qquad (1-\xi^2) ^{3n+1/2} \eta_n (\xi), \qquad  (1-\xi^2) ^{3n-1/2} t_n (\xi)
\ee
are {\it polynomials} in $\xi$. Moreover, we have found that the explicit solution for $R_n(\xi)$ has no singularity at $\xi=1$ and it is holomorphic in the half-plane ${\rm Re}(\xi)>-1$, as expected from general principles. This involves a non-trivial cancellation of poles at $\xi=1$ in the sum of terms (\ref{rss}) giving $R_n(\xi)$, and it can be regarded as a check of the procedure. 

Although we have not studied in detail the structure of the functions $r_n(\xi)$, $\eta_n(\xi)$, some patterns can be easily observed. As we will see in a moment, it is useful to use, instead of the variable $\xi$, the variable $u$, defined by 
\be
\label{uvar}
u={4 \xi \over z}. 
\ee
The dominant terms in $r_n$, $\eta_n$ for $z\rightarrow \infty$ and $u$ fixed seem to have the general structure
\be
\label{rn-an}
\ba
r_n(u,z)&\sim {z^{6n-2} u^{6n-1}\over \left(1 - {z^2 u^2 \over 16} \right)^{3n+{1\over 2}}}   {\pi^{2n} \over (2n)! 2^{14n-2}}, \\
\eta_n(u,z)&\sim -{z^{6n-1} u^{6n+1}\over \left(1 - {z^2 u^2 \over 16} \right)^{3n+{1\over 2}}}   {\pi^{2n} \over (2n)! 2^{14n}}.
\ea
\ee
This has been guessed by looking at the first orders in the explicit solution, but we have not found an analytic proof. As we will see in the next section, however, the above structure for $r_n$ implies the conjecture for the form of the one-membrane function $a_1(k)$ conjectured in \cite{hmo2}. 

\subsection{The grand potential} 

Our ultimate goal is to compute the WKB expansion of $J^{\rm M2}(\mu, k)$, therefore we have to compute the grand potential. This 
follows from (\ref{grand-k}). We will denote 
\be
J_z \equiv {\partial J  \over \partial z}. 
\ee
It is convenient to split this quantity w.r.t. the parity of $z$, 
\be
 {\partial J  \over \partial z}=J_z^+ + J_z^-, 
 \ee
 where
 \be
 J_z^{\pm}={1\over 4 \pi k } \int_{-\infty}^{\infty} \rd x\, R_\pm(x).
 \ee
 After changing variables from $x$ to 
 \be
 u= {4  \xi \over z}={\rm sech} \left( {x \over 2}\right), 
 \ee
 as defined in (\ref{uvar}), we find 
\be
k J_z^{\pm}= {1\over \pi}  \int_0^1 {\rd u \over u}  \, {1\over {\sqrt{1- u^2}}} R_{\pm}\left( z u /4\right). 
\ee
The WKB expansion of $R_\pm (x)$ leads to the WKB expansion of the grand potential considered in \cite{mp}, 
\be
J_z (\mu, k) =\sum_{n \ge 0} J_{z,n} (\mu) k^{2n-1}. 
\ee
The integrals appearing in the calculation of $J_{z,n}$ can be evaluated in terms of generalized hypergeometric functions. For $J_{z,n}^+$, the answer involves the functions $_2F_1$, while for $J_{z,n}^-$ we also 
find the functions $_3F_2$. The result can then be expanded at large $z$, and from this expansion, together with (\ref{gen-M2}), one reads the small $k$ expansion of $a_\ell (k)$, $b_\ell(k)$ and $c_\ell (k)$, for any $\ell$. In the Appendix we collect some useful results on hypergeometric integration which are needed in these calculations, as well as the resulting small $k$ expansion of these coefficients for $\ell=1,2,3$. 

As a simple example of this procedure, let us calculate the leading order correction to the grand potential, $J_{z,0}$. We have
\be
J_{z,0}^+= {1\over \pi} \int_{0}^{1} {\rd u \over {\sqrt{1- u^2}} {\sqrt{1-z^{2} u^{2}/16}}}={1\over 2\pi} K \left( {z^2 \over 16} \right), 
\ee
and 
\be
\ba
J_{z,0}^-&= {1\over \pi^2} \int_{0}^{1} \rd u \sin^{-1} \left( {u z \over 4} \right) {1\over {\sqrt{1- u^2}} {\sqrt{1-z^{2} u^{2}/16}}}=-{z  \over 4 \pi^2}  {~}_3F_2\left(1,1,1;\frac{3}{2},\frac{3}{2};\frac{z^2}{16}\right), 
\ea
\ee
where we used (\ref{fsin}) and (\ref{hyperint}). The above results for $J^\pm_{z,0}$ agree with the calculation in \cite{mp}. We have also checked that the results for $J_{z,1}$ and $J_{z,2}$ agree with the results in \cite{mp}. 

It is interesting to notice that, in order to read off the coefficients $a_\ell(k)$, $b_\ell(k)$, it is enough to calculate $J_z^+$: by using the structural result (\ref{gen-M2}) and assuming that the branch cut of the log is along the positive real axis, we find that 
\be
\label{branch-cut}
 J_z^{{\rm M2},+}=\sum_{\ell\ge 1}  \left[  \left( 2 \pi \ri \log z - \pi^2 \right)  \ell a_\ell(k) + \pi \ri \left(\ell b_\ell(k) - a_\ell(k) \right)  \right] z^{-2\ell-1}.
\ee
It is easy to verify that the leading order term in the expansion of (\ref{branch-cut}) 
at large $z$ comes from integrating the leading term of $r_n(u,z)$ in 
(\ref{rn-an}). By using (\ref{easy-int-2}) we find
\be
\int_0^1 {\rd u \over {\sqrt{1 -u^2}}} {u^{6n-2} \over \left( 1-{z^2 u^2 \over 16} \right)^{3n+{1\over 2}}}= {1\over 2} {\Gamma(3n-1/2) 
\Gamma(1/2) \over \Gamma(3n)} {~}_2F_1\left( 3n-{1\over 2}, 3n+{1\over 2}, 3n; {z^2 \over 16}\right). 
\ee
At large $z$ we have the logarithmic behavior, 
\be
_2F_1\left( 3n-{1\over 2}, 3n+{1\over 2}, 3n; {z^2 \over 16}\right)= \ri  (-1)^n 2^{12n + 2}  {\Gamma(3n) \over \Gamma(-1/2) \Gamma(3n-1/2)} z^{-6n-1} \log(-z^2) +\cdots
\ee
If we put everything together and we compare the result with (\ref{branch-cut}), we find that the ansatz (\ref{rn-an}) leads to the following perturbative expansion for the coefficient $a_1(k)$:
\be
a_1(k) =- {4\over \pi^2  k } \sum_{n\ge 0} {(-1)^n \over (2n)!} \left( {\pi k \over 2} \right)^{2n} = -{4\over \pi^2 k} \cos\left( {\pi k \over 2}\right). 
\ee
This is precisely the result conjectured in \cite{hmo2} for this coefficient.

 \sectiono{Predictions for membrane instantons}

\subsection{The HMO cancellation mechanism}
The Gopakumar--Vafa representation (\ref{gvone}) of $J^{\rm WS}(\mu, k)$ shows that it has 
double poles at all integer values of $k$. Since the original matrix integral is not singular for any value of $k$, there must be some way of canceling these divergences. The proposal of HMO in \cite{hmo2} is that, in the total non-perturbative grand potential, and order by order in ${\rm e}^{-\mu}$, there should be no divergences. This means that, in the sum (\ref{bound-state}) 
over all bound states $(\ell,m)$ which contribute to a given order in ${\rm e}^{-\mu}$, singularities must cancel. This is the HMO cancellation mechanism.  

In general, since the contribution of generic bound states is not known, it is difficult to verify this cancellation mechanism in detail. 
However, some non-trivial information can be obtained by looking at low orders in the expansion. The case studied in detail in \cite{hmo2} is the case of one-membrane instanton, i.e. of contributions to the grand potential which go like ${\rm e}^{-2\mu}$. Let us assume that $k$ is an even, positive integer. Then, the only contributions to the term of order ${\rm e}^{-2\mu}$ come from 
\be
(\ell,m)=(1,0) \quad {\text{and}} \quad (\ell,m)=(0,k/2), 
\ee
i.e. they involve only worldsheet instantons and membrane instantons. In this case, the HMO cancellation mechanism can be studied in detail. The worldsheet instanton poles at $k=2m$ are of the form
\be
\label{one-inst}
d_m(k)\re^{-4 m \mu/k}=(-1)^{m-1} \left[ { 4 m \over  \pi^2 (k-2m)^2} + {4 (\mu + 1) \over \pi^2 (k-2m)} + {2 \mu^2 + 2 \mu +1 \over m \pi^2}+ w^{(m)}\right] \re^{-2 \mu} +\cdots
\ee
where $w^{(m)}$ can be calculated from the Gopakumar--Vafa invariants of local $\IP^1\times \IP^1$. According to the HMO mechanism, the poles have to be cancelled by similar poles in the membrane instanton contribution. Using these constraints, as well as the first three terms of the WKB expansion calculated in \cite{mp}, HMO were able to propose an ansatz for the values of the coefficients $a_1(k)$, $b_1(k)$, $c_1(k)$ appearing in (\ref{gen-M2}), 
\be
\label{one-ansatz}
\ba
a_1(k)&= -{4 \over \pi^2 k} \cos\left( {\pi k \over 2}\right), \\
b_1(k)&=  \frac{2}{\pi}\cos^2\left(\frac{\pi k}{2}\right)\csc\left(\frac{\pi k}{2}\right),\\
c_1(k)&=\left[ -{2\over 3k} + {5 k \over 12} + {k \over 2} \csc^2 \left( {\pi k \over 2}\right)+ {1\over \pi} \cot \left( {\pi k \over 2}\right) \right]  \cos\left( {\pi k \over 2}\right). 
\ea
\ee
One can check that these expressions have the right singularity structure to cancel the divergences in (\ref{one-inst}): near $k=2m$, one has
\be
\label{bc-one}
\ba
b_1(k) &= -{4 (-1)^{m-1}  \over \pi^2 (k-2m)} +\CO(k-2m),\\
c_1(k) &= (-1)^{m-1} \left[ -{4 m  \over \pi^2 (k-2m)^2} - {4 \over \pi^2 (k-2m)}+{1\over 3m}- {2m \over 3} \right]+\CO(k-2m).
\ea
\ee
Once the membrane instanton and the worldsheet instanton contributions are added for even integer $k=2m$, the poles cancel and we find a finite contribution 
\be
(-1)^{m-1} \left[ {4 \mu^2 + 2 \mu + 1 \over m \pi^2} + w^{(m)}+{1\over 3m}- {2m \over 3} \right] \re^{-2\mu}.
\ee
This result reproduces a numerical calculation of the terms of order $\re^{-2\mu}$ in $J(\mu, k)$ for $k=2,4,6$, and confirms the ansatz (\ref{one-ansatz}).  

The conjecture (\ref{one-ansatz}) shows that there is a simple, hidden structure in the contribution of membrane instantons to the grand potential. Thanks to our semiclassical TBA expansion, 
we can now compute the series expansion of these coefficients to higher order in $k$, as listed in the Appendix. Our results confirm the expressions in (\ref{one-ansatz}). In the case of $a_1(k)$, as we showed in the last section, this follows from the ansatz (\ref{rn-an}).

\subsection{The second membrane instanton}

In order to provide further tests of the HMO cancellation mechanism with the current techniques, we have to look at exponentially small terms which only receive contributions from 
states of the form $(\ell,0)$ and of the form $(0,m)$. As we saw in the previous section, following \cite{hmo2}, this is what happens for the terms $\sim \re^{-2 \mu}$ when $k$ is even. Similarly, when $k$ is {\it odd}, the exponentially small term ${\rm e}^{-4 \mu}$ has contributions only from the states
\be
(\ell,m)=(2,0) \quad {\text{and}} \quad (\ell,m)=(0,k),
\ee
and it involves the two-membrane instanton. This case was only partially analyzed in \cite{hmo2}. They noted that the contribution of the 
worldsheet instanton for $k=m$ odd is singular, and it has the behavior 
\be
\label{two-poles}
d_m (k) \re^{-4 m \mu/k}= \left[ { m \over 4 \pi^2 (k-m)^2} + {2\mu + 1 \over 2 \pi^2 (k-m)} + {2 \mu^2 + \mu +1/4 \over m \pi^2}+ v^{(m)}\right] \re^{-4 \mu} +\cdots
\ee
The $v^{(m)}$ can be calculated from the Gopakumar--Vafa invariants and they are given, for $m=1,3$, by  
\be
v^{(1)}= {1\over 3}, \qquad v^{(3)}= {37 \over 9}.
\ee
The poles appearing in (\ref{two-poles}) were also conjectured in \cite{hmo2} to be cancelled by similar poles in the membrane instanton contribution. 
Using this cancellation mechanism, as well as our data for the WKB expansion, we propose the following exact expressions for the coefficients of the second membrane instanton: 
\be
\ba
a_2(k)&=-{1\over \pi^2 k } \left(8 +10 \cos\left( \pi k\right) \right),\\
b_2(k)&={4 \over \pi^2 k} \left( 1 + \cos\left( \pi k\right)  \right) + {1\over 2 \pi }\csc \left( \pi k\right) \left( 17 + 24 \cos\left( \pi k\right) + 9 
\cos\left(2 \pi k\right)  \right), \\
c_2(k)&=-\frac{4}{3 k}-\frac{5 \cos (\pi  k)}{3 k}+k \left(\frac{49}{24} \cos (\pi  k)-\frac{7}{6}\right)+\frac{\cot (\pi  k)}{\pi }+5 k \csc ^2(\pi
    k)\\
    &+\cos (\pi  k) \left(\frac{5 \cot (\pi  k)}{4 \pi }+\frac{21}{4} k \csc ^2(\pi  k)\right).
\ea
\ee
It is easy to check that the above expressions reproduce the WKB expansions presented in the Appendix. They also cancel the poles in (\ref{two-poles}). Indeed, one finds that, near $k=m$ odd, 
\be 
\ba
b_2(k)&= -{1\over  \pi^2 (k-m)} +\CO(k-m), \\
c_2(k)&=-{m \over 4 \pi^2 (k-m)^2} -{1\over 2 \pi^2 (k-m)} +  {1\over 3m}-{2m \over 3}+ \CO(k-m).
\ea
\ee
The finite piece, after adding both contributions, is 
\be
 \left[ {4 \mu^2  + \mu +1/4\over m \pi^2} + v^{(m)}+ {1\over 3m}-{2m \over 3} \right]\re^{-4 \mu}. 
 \ee
 For $m=1,3$ we find
 \be
  \left[ {4 \mu^2  + \mu +1/4\over  \pi^2} \right]\re^{-4 \mu}\quad {\text {and}}  \quad  \left[ {4 \mu^2  + \mu +1/4\over 3 \pi^2} + {20 \over 9} \right]\re^{-4 \mu},
  \ee
respectively, which correctly reproduce the terms of order $\re^{-4 \mu} $ in the expression for $J(\mu,k)$ with $k=1,3$ presented in \cite{hmo2}. Notice that, although the coefficient $c_2(k)$ is very different from the coefficient $c_1(k)$, their finite parts near the relevant poles have the same form.

\subsection{Higher membrane instantons}

The HMO cancellation mechanism can be studied in detail for one-membrane instantons and two-membrane instantons. The contribution of higher membrane instantons mixes with the contribution of generic bound states and it is difficult to extract information about the former without a more detailed knowledge of the latter. In particular, 
we don't have enough information about the singularity structure of the membrane coefficients as a function of $k$, and so far the only available information is contained in the WKB expansions. 

In the case of the $a_\ell (k)$ coefficients, however, it is easy to fit these WKB expansions to a sum of trigonometric functions. We find that, for odd (even) instanton number, the $a_\ell(k)$ 
are given by a sum of cosines whose argument is an odd (respectively, even) integer times $\pi k/2$. Their explicit expressions, up to instanton number $\ell=5$, are
\be
\ba
a_3(k)&=-{1\over \pi^2 k }  \left(88 \cos\left( {\pi k \over 2}\right) +{124 \over 3} \cos\left( {3 \pi k \over 2}\right) +4 \cos\left( {5 \pi k \over 2}\right)  \right),\\
 a_4(k)&= -\frac{1}{\pi^2 k }(364+560\cos(\pi k)+245\cos(2\pi k)+48\cos(3\pi k)+8\cos(4\pi k)), \\
 a_5(k) &= -\frac{1}{\pi^2 k }\left(6080\cos\left(\frac{\pi k}{2}\right) + 4100\cos\left(\frac{3\pi k}{2}\right) + {9104 \over 5} \cos\left(\frac{5\pi k}{2}\right) + 536\cos\left(\frac{7\pi k}{2}\right) \right.\\
  &\left.+ 136\cos\left(\frac{9\pi k}{2}\right) + 24\cos\left(\frac{11\pi k}{2}\right) + 4\cos\left(\frac{13\pi k}{2}\right)\right). 
  \ea
\ee
Finding a natural expression which fits our data for the expansion of the $b_\ell(k)$, $c_\ell(k)$ with $\ell\ge 3$ is more challenging, and further information (including more data points in the WKB expansion) is probably needed. 

\sectiono{Conclusions and open problems}

In this paper we have developed a semiclassical approach to the TBA equations of \cite{zamo}. When applied to ABJM theory, this reproduces in a more efficient way the WKB approach developed in \cite{mp}. Our results confirm the conjecture made in \cite{hmo2} for the one-instanton contribution. We have also proposed analytic expressions at finite $k$ for the membrane instantons at order two, which are in perfect accord with the HMO cancellation mechanism. In addition, we have obtained some conjectural results at higher instanton number for the coefficients $a_\ell (k)$. 

It is obvious from the results in this paper and its predecessors that there is a new and rich story concerning non-perturbative corrections in the M-theory dual to ABJM theory. Corrections to the partition function coming from intrinsic M-theory objects can be now computed in detail, and sometimes we can even guess their exact expression for arbitrary $k$. Moreover, the HMO cancellation mechanism shows very clearly that the genus expansion of type IIA string theory, although it can be resummed with a Gopakumar--Vafa representation, is essentially incomplete at strong coupling: only when the contribution of membrane instantons is taken into account do we find a finite answer for integer values of $k$ (precisely the values for which we believe that the theory is defined non-perturbatively). 

On the other hand, it is fair to say that we are only in a preliminary exploratory period of all these non-perturbative phenomena. Although the ABJM matrix integral and the 
TBA system contain detailed information about these instanton effects, it is not obvious how to extract it. The techniques proposed in \cite{py,hmo,hmo2} work for finite $k$, but they are based on a small $z$ expansion which corresponds to small $N$. In this paper we have proposed a WKB expansion which is only valid for small $k$, but provides analytic large $N$ results. Clearly, both approaches are  insufficient, and they should be combined: what we need is the large $\mu$ expansion of the grand potential at finite $k$. This will very likely require a clever analysis of the TBA system which circumvents the problem with the Bose--Einstein condensate singularity pointed out in section 3. 

A method leading to results at large $N$ and finite $k$ would also determine the functions $f_{\ell,m}(\mu, k)$ appearing in (\ref{bound-state}) for general bound states (maybe order by order in ${\rm e}^{-\mu}$). In fact, it seems to us that the most important open problem is finding an approach to calculate the contribution of bound states, about which nothing is known. Even a WKB approach to the problem would be useful. It was suggested in \cite{mp} that worldsheet instantons appear as quantum-mechanical instantons in the Fermi gas, and bound states might then appear as exponentially small corrections in $\mu$ to a general quantum-mechanical instanton amplitude. This approach might give a first handle on this problem. Another possibility is to look for trans-series solutions of the difference equations (\ref{tba-k}), similar to what was done in \cite{mmnp} for the difference equations arising in matrix models.

In addition to these general issues, there are a multitude of more concrete questions that come to our mind when we look at the conjectural expressions for the membrane instantons. Their contribution involves trigonometric functions whose coefficients are rational numbers and, very often, integer numbers. Is there some sort of Gopakumar--Vafa formula for them? Is there some duality between membrane instantons and worldsheet instantons? Can we compute some of these membrane instanton corrections directly in M-theory, by adapting for example the framework of \cite{hm}?

Finally, it was pointed out in \cite{mp,mytalk} that any non-perturbative result in the context of the Fermi gas of ABJM theory can be interpreted as a non-perturbative result for topological strings in local $\IP^1 \times \IP^1$. In particular, the grand potential studied in this paper can be interpreted as the topological string free energy at large radius, and the membrane instantons computed here lead to corrections to this free energy of order $\CO({\rm e}^{-1/g_s})$. Is there an interpretation for these effects directly in topological string theory? Are there ``topological" membranes which lead to this sort of effect? 

As Bertolt Brecht would put it, ``so many stories, so many questions."

\section*{Acknowledgements}
We would like to thank Albrecht Klemm for useful 
conversations. This work is supported in part by the Fonds National Suisse, subsidies 200020-126817 and 
200020-137523. 

\appendix
\sectiono{Hypergeometric integrals}
In the calculation of the grand potential one finds two types of integrals. For $J_z^+$, we have integrals of the form 
\be
\label{easy-int}
\int_0^1 {\rd u \over {\sqrt{1 -u^2}}} {u^m \over \left( 1-{z^2 u^2 \over 16} \right)^{3s+{1\over 2}}}.  
\ee
After the change of variable 
\be
\label{change-var}
u^2=t, \quad z^2/16= Z
\ee
they can be written as
\be
\label{easy-int-2}
{1\over 2} \int_0^1 \rd t \, (1-t)^{-1/2} t^{m-1\over 2} (1- Z t)^{-3 s -1/2}={\Gamma\left( {m + 1\over 2} \right) \Gamma\left({1\over 2} \right) 
\over 2 \Gamma\left( {m\over 2} +1 \right)} {}_2F_1\left( 3s+{1\over 2}, {m+1\over 2}; 1+{m \over 2} ; Z\right). 
\ee

For $J_z^-$, we have to do integrals of the form 
\be
\int_0^1 {\rd u \over {\sqrt{1 -u^2}}} {u^{m-1} \sin^{-1} \left( {u z \over 4} \right) \over \left( 1-{z^2 u^2 \over 16} \right)^{3s+{1\over 2}}}.
\ee
Using that
\be
\label{fsin}
{1\over x} {\sin^{-1}(x) \over {\sqrt{1 -x^2}}} ={}_2F_1\left( 1, 1, {3\over 2}; x^2\right) 
\ee
we can write the above integral as
\be
\label{hard-int}
{z \over 4} \int_0^1 {\rd u \over {\sqrt{1 -u^2}}} u^{m} \left( 1-{z^2 u^2 \over 16} \right)^{-3s} {}_2F_1\left( 1, 1; {3\over 2}; { z^2 u^2 \over 16} \right). 
\ee
To calculate this integral, we need two ingredients. First of all, we use the identity 132 in p. 436 of \cite{prudnikov}, with $n=1$ (we correct a minor misprint in the statement of the identity):
\be
\ba
& (1-x)^{-3s}  {}_2 F_1\left( 1, 1, {3\over 2}; x\right) \\ 
& \qquad \qquad = {(1)_{3 s} \over \left(\frac{1}{2}\right)_{3 s}} 
 \left[  _2F_1\left(1,3 s+1;\frac{3}{2};x\right)-{1\over 6 s} \sum_{k=0}^{3s-1} {(1/2- 3s)_k \over (1-3s)_k} (1-x)^{-k-1} \right].
 \ea
 \ee
 After using this identiy, and the change of variables (\ref{change-var}), the integral (\ref{hard-int}) can be reduced to an integral of the form 
 \be
  \label{hyperint}
 \ba
& \int_0^1 \rd t \, t^{a_{p+1}-1} (1-t)^{b_{q+1} - a_{p+1} -1}  ~_p F_q (a_1, \cdots, a_p; b_1, \cdots, b_q; Z t) \\
& = 
 {\Gamma(a_{p+1}) \Gamma(b_{q+1} - a_{p+1} ) \over \Gamma (b_{q+1})} ~_{p+1} F_{q+1} (a_1, \cdots, a_p, a_{p+1}; b_1, \cdots, b_q, b_{q+1}; Z )
 \ea
 \ee
 with $p=2$, $q=1$, as well as to a sum of integrals of the form (\ref{easy-int}) with half-integer $s$. 
 
The asymptotic behavior as $z \gg 1$ of the hypergeometric functions involved in these expressions can be easily found by using for example the Barnes representation. For the function $_2F_1$, it is given by (see for example \cite{lebedev}, eq. (9.7.7)):
\be
\ba
&_2F_1(\alpha, \alpha+n; \gamma; z) =   {\Gamma(\gamma) (-z)^{-\alpha}\over \Gamma (\gamma-\alpha) \Gamma(\alpha +n)
}  \sum_{i=0}^{n-1} {(n - i - 1)! (\alpha)_i (1 - \gamma + \alpha)_i  \over i! } (-z)^{-i}\\
&+ {\Gamma(\gamma)(-z)^{-\alpha} \over \Gamma (\gamma-\alpha-n) \Gamma(\alpha)
}  \sum_{i=0}^{\infty} { (\alpha+n)_i (1 - \gamma + \alpha+n)_i  \over i! (n+i)!} \bigl[ \psi(i+1) + \psi(n+i+1) \\&
\qquad \qquad  -\psi(\alpha+ n+i)- \psi(\gamma-\alpha-n-i) + \log(-z) \bigr] z^{-i-n}.
\ea
\ee

\section{WKB expansions}

Here we list the WKB expansion of the membrane instanton coefficients $a_\ell(k)$, $b_\ell(k)$, $c_\ell(k)$ for $\ell=1,2,3$. 
\be
\ba
k \, a_1(k)&=-\frac{4}{\pi ^2}+\frac{k^2}{2}-\frac{\pi ^2 k^4}{96}+\frac{\pi ^4
   k^6}{11520}-\frac{\pi ^6 k^8}{2580480}+\frac{\pi ^8
   k^{10}}{928972800}-\frac{\pi ^{10} k^{12}}{490497638400}\\
   &+\frac{\pi
   ^{12} k^{14}}{357082280755200}-\frac{\pi ^{14}
   k^{16}}{342798989524992000}+\CO\left(k^{18}\right),
   \ea
   \ee
   \be
   \ba
k \, a_2 (k)&=   -\frac{18}{\pi ^2}+5 k^2-\frac{5 \pi ^2 k^4}{12}+\frac{\pi ^4
   k^6}{72}-\frac{\pi ^6 k^8}{4032}+{\pi ^8
   k^{10} \over 362880}-\frac{\pi ^{10} k^{12}}{47900160}+\frac{\pi ^{12}
   k^{14}}{8717829120} \\  
   &-\frac{\pi ^{14}
   k^{16}}{2092278988800}+\CO\left(k^{18}\right),
    \ea
   \ee
     \be
   \ba
k \, a_3 (k)&=    -\frac{400}{3 \pi ^2}+70 k^2-\frac{371 \pi ^2 k^4}{24}+\frac{1159 \pi ^4
   k^6}{576}-\frac{16373 \pi ^6 k^8}{92160}+\frac{74113 \pi ^8
   k^{10}}{6635520}-\frac{62408051 \pi ^{10}
   k^{12}}{122624409600}\\
   &+\frac{43949569 \pi ^{12}
   k^{14}}{2550587719680}-\frac{5465453813 \pi ^{14}
   k^{16}}{12242821054464000}+\CO\left(k^{18}\right).
   \ea
   \ee
\be
\ba
k \, b_1(k)&=\frac{4}{\pi ^2}-\frac{5 k^2}{6}+\frac{67 \pi ^2 k^4}{1440}-\frac{19 \pi
   ^4 k^6}{48384}+\frac{247 \pi ^6 k^8}{38707200}+\frac{89 \pi ^8
   k^{10}}{1226244096}+\frac{1430857 \pi ^{10}
   k^{12}}{669529276416000}\\ &+\frac{1637 \pi ^{12}
   k^{14}}{30607052636160}+\frac{118522319 \pi ^{14}
   k^{16}}{87413742328872960000}+\CO\left(k^{18}\right),
    \ea
   \ee
     \be
   \ba
   k\, b_2(k)&=\frac{33}{\pi ^2}-\frac{77 k^2}{6}+\frac{119 \pi ^2 k^4}{72}-\frac{1199
   \pi ^4 k^6}{15120}+\frac{251 \pi ^6 k^8}{120960}-\frac{607 \pi ^8
   k^{10}}{119750400}+\frac{311813 \pi ^{10}
   k^{12}}{130767436800}\\ & +\frac{56863 \pi ^{12}
   k^{14}}{261534873600}+\frac{4740503 \pi ^{14}
   k^{16}}{213412456857600}+\CO\left(k^{18}\right), \ea
   \ee
     \be
   \ba
  k\, b_3(k)&= \frac{2560}{9 \pi ^2}-188 k^2+\frac{4769 \pi ^2 k^4}{90}-\frac{258689 \pi
   ^4 k^6}{30240}+\frac{159091 \pi ^6 k^8}{172800}-\frac{262188523 \pi ^8
   k^{10}}{3832012800}\\ & +\frac{160190711489 \pi ^{10}
   k^{12}}{41845579776000}-\frac{15646173899 \pi ^{12}
   k^{14}}{133905855283200}+\frac{14718859878607 \pi ^{14}
   k^{16}}{1365839723888640000}+\CO\left(k^{18}\right).
   \ea
   \ee
   \be
   \ba
   k \, c_1(k)&=\left(\frac{4}{\pi ^2}-\frac{2}{3}\right)+\frac{1}{12} \left(\pi
   ^2-1\right) k^2+\left(-\frac{13 \pi ^2}{360}-\frac{\pi ^4}{576}\right)
   k^4+\left(\frac{55 \pi ^4}{96768}+\frac{\pi ^6}{69120}\right)
   k^6\\  & +\left(-\frac{671 \pi ^6}{38707200}-\frac{\pi ^8}{15482880}\right)
   k^8+\left(\frac{\pi ^{10}}{5573836800}-\frac{3659 \pi
   ^8}{12262440960}\right) k^{10}\\ & +\left(-\frac{713927 \pi
   ^{10}}{66952927641600}-\frac{\pi ^{12}}{2942985830400}\right)
   k^{12}+\CO\left(k^{14}\right), \ea
   \ee
     \be
   \ba
 k \, c_2(k)&=  \left(\frac{25}{2 \pi ^2}-3\right)+\left(\frac{7}{24}+\frac{5 \pi
   ^2}{6}\right) k^2+\left(-\frac{47 \pi ^2}{60}-\frac{5 \pi
   ^4}{72}\right) k^4+\left(\frac{641 \pi ^4}{8640}+\frac{\pi
   ^6}{432}\right) k^6\\  &+\left(-\frac{3443 \pi ^6}{1209600}-\frac{\pi
   ^8}{24192}\right) k^8+\left(\frac{29 \pi ^8}{4561920}+\frac{\pi
   ^{10}}{2177280}\right) k^{10}\\  &+\left(-\frac{26897 \pi
   ^{10}}{4572288000}-\frac{\pi ^{12}}{287400960}\right)
   k^{12}+\CO\left(k^{14}\right), \ea
   \ee
     \be
   \ba
 k \, c_3(k)&=   \left(\frac{1642}{27 \pi
   ^2}-\frac{200}{9}\right)+\left(\frac{437}{36}+\frac{35 \pi
   ^2}{3}\right) k^2+\left(-\frac{10987 \pi ^2}{576}-\frac{371 \pi
   ^4}{144}\right) k^4\\  &+\left(\frac{182071 \pi ^4}{32256}+\frac{1159 \pi
   ^6}{3456}\right) k^6+\left(-\frac{3294323 \pi ^6}{3686400}-\frac{16373
   \pi ^8}{552960}\right) k^8\\  &+\left(\frac{769550747 \pi
   ^8}{8758886400}+\frac{74113 \pi ^{10}}{39813120}\right)
   k^{10}+\left(-\frac{43661410369 \pi
   ^{10}}{7084965888000}-\frac{62408051 \pi ^{12}}{735746457600}\right)
   k^{12}+\CO\left(k^{14}\right).
   \ea
   \ee

\end{document}